\documentclass{aa}  

\usepackage[dvipsnames]{xcolor} 
\usepackage[para,online,flushleft]{threeparttable}
\usepackage{siunitx}
\usepackage{CJKutf8}
\usepackage{enumitem}
\usepackage{afterpage}
\usepackage{graphicx}
\usepackage{caption}
\usepackage{subcaption,comment}
\usepackage{txfonts}
\usepackage{hyperref}
 \hypersetup{
     colorlinks=true,
     linkcolor=blue,
     citecolor=blue, 
     urlcolor=blue,
     }

\usepackage{orcidlink}

\begin{document}

   \title{A MeerKAT view of the double pulsar eclipses}
  \subtitle{Geodetic precession of pulsar B and system geometry}

   \author{
          M.~E. Lower \inst{\ref{csiro}}
          \and 
          M. Kramer \inst{\ref{mpifr},\ref{jbca}}
          \and
          R.~M. Shannon \inst{\ref{swin}, \ref{ozgrav}}
          \and
          R.~P. Breton \inst{\ref{jbca}}
          \and
          N. Wex \inst{\ref{mpifr}}
          \and
          S. Johnston \inst{\ref{csiro}}
          \and
          M. Bailes \inst{\ref{swin}, \ref{ozgrav}}
          \and
          S. Buchner \inst{\ref{sarao}}
          \and
          H.~Hu~(胡奂晨) \inst{\ref{mpifr}}
          \and
          V. Venkatraman Krishnan \inst{\ref{mpifr}}
          \and
          V.~A. Blackmon \inst{\ref{uva}}
          \and
          F. Camilo \inst{\ref{sarao}}
          \and
          D.~J. Champion \inst{\ref{mpifr}}
          \and
          P.~C.~C. Freire \inst{\ref{mpifr}}
          \and\\
          M. Geyer \inst{\ref{uct}}
          \and
          A. Karastergiou \inst{\ref{oxford}, \ref{rhodes}}
          \and
          J.~van~Leeuwen \inst{\ref{astron}}
          \and
          M.~A. McLaughlin \inst{\ref{uva}}
          \and
          D.~J. Reardon \inst{\ref{swin}, \ref{ozgrav}}
          \and
          I.~H. Stairs \inst{\ref{ubc}}
          }

   \institute{Australia Telescope National Facility, CSIRO, Space and Astronomy, PO Box 76, Epping, NSW 1710, Australia\label{csiro}
              \and
              Max-Planck-Institut f\"{u}r Radioastronomie, Auf dem H\"{u}gel 69, D-53121 Bonn, Germany\label{mpifr}
              \and
              Jodrell Bank Centre for Astrophysics, The University of Manchester, Oxford Road, Manchester M13 9PL, United Kingdom\label{jbca} 
              \and    
              Centre for Astrophysics and Supercomputing, Swinburne University of Technology, PO Box 218, Hawthorn, VIC 3122, Australia\label{swin} 
              \and
              ARC Centre of Excellence for Gravitational Wave Discovery (OzGrav) \label{ozgrav}
              \and
              South African Radio Astronomy Observatory, 2 Fir Street, Black River Park, Observatory 7925, South Africa\label{sarao}
              \and
              Department of Physics and Astronomy, West Virginia University, Morgantown, WV 26501\label{uva}
              \and
              Department of Mathematics and Applied Mathematics, University of Cape Town, Rondebosch, 8001, South Africa\label{uct}
              \and
              Department of Astrophysics, University of Oxford, Denys Wilkinson Building, Keble Road, Oxford OX1 3RH, UK\label{oxford}
              \and
              Department of Physics and Electronics, Rhodes University, PO Box 94, Grahamstown 6140, South Africa\label{rhodes}
              \and
              ASTRON, the Netherlands Institute for Radio Astronomy, Oude Hoogeveensedijk 4,7991 PD Dwingeloo, The Netherlands\label{astron}
              \and
              Dept. of Physics and Astronomy, University of British Columbia, 6224 Agricultural Road, Vancouver, BC V6T 1Z1 Canada\label{ubc}\\\\ 
              \email{marcus.lower@csiro.au}
             }

\date{Received MM dd, yyyy; accepted MM dd, yyyy}

\abstract{
The double pulsar system, PSR~J0737$-$3039A/B, consists of two neutron stars bound together in a highly relativistic orbit that is viewed nearly edge-on from the Earth.
This alignment results in brief radio eclipses of the fast-rotating pulsar A when it passes behind the toroidal magnetosphere of the slow-rotating pulsar B. 
The morphology of these eclipses is strongly dependent on the geometric orientation and rotation phase of pulsar B, and their time evolution can be used to constrain the geodetic precession rate of the pulsar.
We demonstrate a Bayesian inference framework for modelling high-sensitivity eclipse light curves obtained with MeerKAT between 2019 and 2023.
Using a hierarchical inference approach, we obtained a precession rate of $\Omega_{\rm SO}^{\rm B} = {5.16^{\circ}}^{+0.32^{\circ}}_{-0.34^{\circ}}$\,yr$^{-1}$ (68\% confidence intervals) for pulsar B, consistent with predictions from general relativity to a relative uncertainty of 6.5\%.
This updated measurement provides a 6.1\% test of relativistic spin-orbit coupling in the strong-field regime.
We show that a simultaneous fit to all of our observed eclipses can in principle return a $\sim$1.5\% test of spin-orbit coupling. 
However, systematic effects introduced by the current geometric orientation of pulsar B along with inconsistencies between the observed and predicted eclipse light curves result in difficult to quantify uncertainties when using this approach.
Assuming the validity of general relativity, we definitively show that the spin axis of pulsar B is misaligned from the total angular momentum vector by $40.6^{\circ} \pm 0.1^{\circ}$ and that the orbit of the system is inclined by approximately $90.5^{\circ}$ from the direction of our line of sight. 
Our measured geometry for pulsar B suggests the largely empty emission cone contains an elongated horseshoe-shaped beam centred on the magnetic axis, and that it may not be re-detected as a radio pulsar until early 2035.
}

\begin{CJK*}{UTF8}{gkai}
\maketitle
\end{CJK*}

\section{Introduction}

PSR~J0737$-$3039A/B is a highly relativistic double neutron star binary with a short, 2.45\,hr mildly eccentric ($e = 0.088$) orbit \citep{Burgay2003}. 
Uniquely, both neutron stars have been detected as radio pulsars (hereafter referred to as pulsars A and B) with respective spin periods of 22.7\,ms and 2.77\,s \citep{Lyne2004}. 
As a result, the system was dubbed the `double pulsar'.
High-precision timing of the two pulsars resulted in four independent tests of Einstein's general theory of relativity (GR) in the strong-field regime within only 2.7\,yr of its initial discovery \citep{Kramer2006}.
Continued timing of the faster rotating pulsar A enabled a preliminary measurement of Lense-Thirring precession in the system, which was used to place limits on the pulsar's moment of inertia, along with several higher-order relativistic effects \citep{Kramer2021b, Hu2022}.
These detections have made the double pulsar one of the most successful astrophysical laboratories for testing our theories of gravity and the behaviour of matter at super-nuclear densities.

A substantial contributor to our ability to test fundamental physics with this system comes from its remarkably edge-on orientation of the orbital plane, which is inclined at an angle of $i = 89.35^{\circ} \pm 0.05^{\circ}$ (or equivalently, $90.65^{\circ} \pm 0.05^{\circ}$) from our perspective on Earth \citep{Kramer2021b}. 
This chance geometric alignment results in a 30-40\,s long eclipse of pulsar A by the magnetosphere of pulsar B around superior conjunction, offering a unique means to directly probe the plasma environment around an active pulsar \citep{Lyne2004}. 
The duration of these eclipses corresponds to a region of space that is $\sim$1.7$\times 10^{7}$\,m wide, spanning only $\sim$10\% of the light-cylinder radius of an equivalent isolated pulsar with the same rotational properties as pulsar B \citep{Lyne2004, Kaspi2004, Breton2012}. 
This smaller than expected eclipsing region arises from the relativistic wind from pulsar A penetrating deep into the magnetosphere of pulsar B, compressing the `windward' side facing pulsar A and blowing the `leeward' side backwards into a cometary magnetotail \citep{Arons2005}.

High-time resolution observations of the eclipses analysed by \citet{McLaughlin2004} with the 100-m Green Bank Telescope (GBT) revealed the light curve of pulsar A exhibits peaks and troughs in its flux density that are spaced apart by once and half the 2.77\,s rotation period of pulsar B at different eclipse phases. 
This modulation can be almost entirely explained through a simple geometric model, in which the radio pulses undergo synchrotron absorption by the relativistic pair plasma that is confined to the toroidal, closed-field region of pulsar B's magnetosphere \citep{Lyutikov2005b}.
The success of the model provided not only the first direct evidence for a dipole magnetic field geometry around a pulsar, but was later used to track temporal changes seen in a set of eclipses collected with the GBT over 3.9\,yr, resulting in a method of detecting the geodetic precession rate of pulsar B ($\Omega^{\rm B}_{\rm SO}$) and an associated fifth independent test of GR \citep{Breton2008}.
While the effects of geodetic precession have been detected in six other relativistic binary pulsars to date \citep{Kramer1998, Kirsten2014, Fonseca2014, VenkatramanKrishnan2019, vanLeeuwen2015, Cameron2023}, the corresponding precession rate measurements are largely indirect and primarily based on modelling the pulse profile width and polarisation properties. 
Interpretation of these observables can be heavily reliant on the assumed pulsar beam shape, system geometry, and the applicability of the rotating vector model \citep{Radhakrishnan1969a}, all of which can display large deviations from simple models.
\citet{Breton2008} demonstrated the precession of pulsar B has a significant impact on the observed eclipse light curve, which was fitted by adding a simple linear drift in the spin-axis longitude of pulsar B over time when computing the model templates.

In addition to enabling tests of spin-orbit coupling in the strong-field regime, the eclipses allow us to develop an improved picture of the overall geometry of the double pulsar. 
Such measurements provide an important input into the construction of stellar binary population synthesis models \citep[e.g.][]{Compas2022}, improving our understanding of the formation and evolutionary history of such systems \citep{Stairs2006, Kim2015, Tauris2017, Vigna-Gomez2018}, and generating astrophysically motivated priors for analysing gravitational waves from double neutron star mergers \citep{Zhu2020}.
Within months of pulsar B being discovered, it became obvious that the overall shape, intensity and position of the profile within the brief `orbital visibility windows' were evolving with time \citep{Burgay2005}.
This behaviour is linked to our changing line of sight through the emission cone as the pulsar undergoes geodetic precession.
While this ultimately resulted in the disappearance of radio pulses from pulsar B sometime in late-2008 \citep{Perera2010}, it did open the rare opportunity to develop a map of the radio beam.
Similar beam maps have only been obtained for three other pulsars in relativistic binaries to date (B1913$+$16, \citealt{Weisberg2002}; J1141$-$6145, \citealt{Manchester2010}; and J1906$+$0746, \citealt{Desvignes2019}).
Accurate geometric constraints are required when creating such radio-beam maps and predicting when radio pulses from  pulsar B may once again be detectable \citep{Breton2009, Perera2012, Noutsos2020}.

In this work, we present the first results of an eclipse monitoring campaign that utilises the low system temperature and high gain of MeerKAT to capture the subtle time-evolution of the eclipse modulation pattern in the greatest detail yet.
Section~\ref{sec:obs} describes the observations and data processing steps. 
In Section~\ref{sec:methods} we outline our improved approach to modelling the eclipse light curves, and the use of hierarchical inference methods to infer the geodetic precession rate of pulsar B.
The results from our eclipse modelling are detailed in Section~\ref{sec:results}, including an outline of the challenges faced given the current system geometry and model compatibility.
In Section~\ref{sec:discussion} we discuss updates to the tests of GR and constraints on the system geometry that stem from our improved measurement of geodetic precession, implications for the radio beam shape of pulsar B and when we may again detect radio pulses from it.
We summarise the results in Section~\ref{sec:conclusion} and outline potential future directions for modelling of the eclipses.

\section{Observations}
\label{sec:obs}

We have performed monthly monitoring observations of the double pulsar from July 2019 to September 2022 with the Meer Karoo Array Telescope (MeerKAT), a 64-element radio interferometer located in the Northern Cape province of South Africa \citep{Jonas2016}.
Our observations were collected under the Relativistic Binary theme of the MeerTime large science project \citep{Bailes2020, Kramer2021a}. 
Observations at MeerKAT from March 2019 to March 2020 were performed using the 1284\,MHz central frequency L-band receiver system, after which the majority of observations were performed with the 816\,MHz central frequency UHF receivers.
These data were collected using the PTUSE instrument (\citealt{Bailes2020}) and are coherently dedispersed to account for the frequency-dependent delay induced by the passage of pulsar A's radiation through the interstellar medium. 
PTUSE provides 1024 frequency channel filterbank data across the 856\,MHz and 544\,MHz bandwidths of the L-band and UHF receiver fleets respectively, along with $\sim$9\,$\mu$s time sampling, and full Stokes information \citep{Bailes2020}. 

The coherent search-mode data were folded at the predicted rotation period of pulsar A using the {\sc dspsr} software package \citep{vanStraten2011}, where individual single pulses were saved to {\sc psrchive} format archive files \citep{Hotan2004, vanStraten2012}.
Frequency channels that were affected by radio-frequency interference were excised using the {\sc MeerGuard}\footnote{\url{https://github.com/danielreardon/MeerGuard}} pulsar data cleaning package.
We then averaged the data in frequency and binned in time by four rotations of pulsar A for an effective time resolution of $\sim$91\,ms.
Only the total intensity data were analysed in this work. 
Analysis of the polarimetry of pulsar A throughout the eclipses is the subject of a separate work.
We also saved a copy of the data set where frequency-channels that were outside a 961--1088\,MHz subband that is covered by both the UHF and L-band receivers were excised.
This was done to test whether or not frequency dependencies in the eclipse width and depth affect our ability to infer secular changes in the modulation pattern.
The two data sets are distinguished by the {\sc Fullband} and {\sc Subband} designations hereafter.

\section{Methods}\label{sec:methods}

\subsection{Eclipse light-curve extraction}

\begin{figure}[t!]
    \centering
    \includegraphics[width=\linewidth]{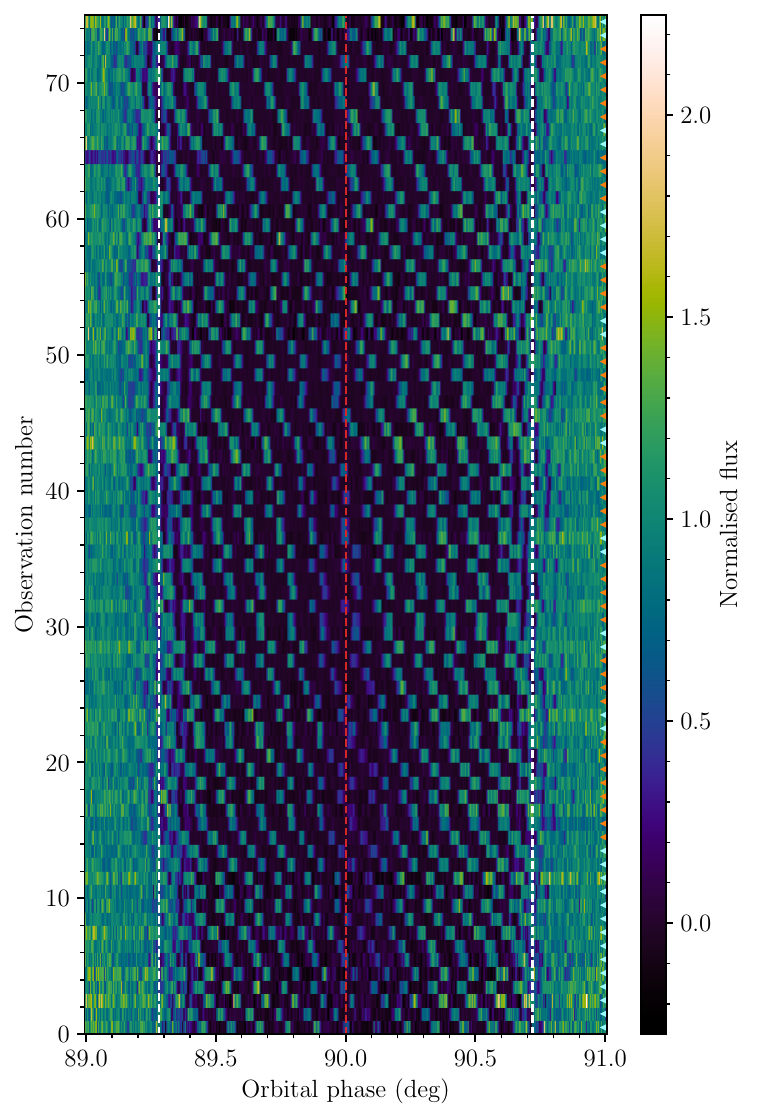}
    \caption{Normalised intensity of pulsar A throughout each of the eclipses recorded by MeerKAT. The red vertical line indicates the orbital phase of superior conjunction. Regions contained within the white lines correspond to where the light curve model best matches the data (see Section~\ref{subsec:limitaions}). Cyan and orange indicators on the right-hand side denote light-curves obtained using the L-band and UHF receivers respectively. We note that the observation numbers on the y-axis do not increase linearly with time.}
    \label{fig:0737_estack}
\end{figure}

To model the total intensity light curve, we first extracted the flux of pulsar A via a matched-filtering process.
This was performed using the {\sc psrflux} tool in {\sc psrchive} \citep{Hotan2004, vanStraten2012}, where a high signal-to-noise template (generated from the integrated pulse profile from many hours of observations) was cross-correlated with a frequency-averaged copy of the total intensity data.
The resulting flux densities were normalised by the median off-eclipse value so the points where pulsar A is unobstructed, have values of order unity.
We used the {\sc pat} tool in {\sc psrchive} to compute topocentric arrival times at MeerKAT for pulsar A, which were converted to equivalent arrival times at the Solar System barycentre using {\sc tempo} and the JPL-DE436\footnote{\url{https://ssd.jpl.nasa.gov/planets/eph_export.html}} Solar System ephemeris.
The barycentred arrival times were then converted to orbital phases using the latest pulsar A timing ephemeris from \citet{Hu2022}\footnote{The orbital phase is obtained from the inverse timing formula (see Section 2.7 of \citealt{Damour1986}). As in \cite{Kramer2021b}, we have performed the inversion numerically.}.
As in \cite{Kramer2021b}, we have taken into account retardation effects due to the motion of pulsar B as the signal from A propagates through the binary system towards Earth.
From here the total-intensity data were then fit using a variation of the recipe outlined in \citet{Breton2008} in order to infer the geometry of pulsar B.

In Figure~\ref{fig:0737_estack} we show the complete sample of eclipse light curves corresponding to the {\sc Fullband} version of the data.
The eclipses are largely dominated by transparency windows, regions where pulsar A is not obscured, that are separated by a full rotation of pulsar B. 
Windows separated by only a half-rotation are restricted to the ingress and egress phases.
This is in contrast to the data analysed by \citet{Breton2008}, where the precession phase of pulsar B meant that the transition from the half to full rotational separation occurred closer to superior conjunction.

\begin{figure}[t!]
    \centering
    \includegraphics[width=\linewidth]{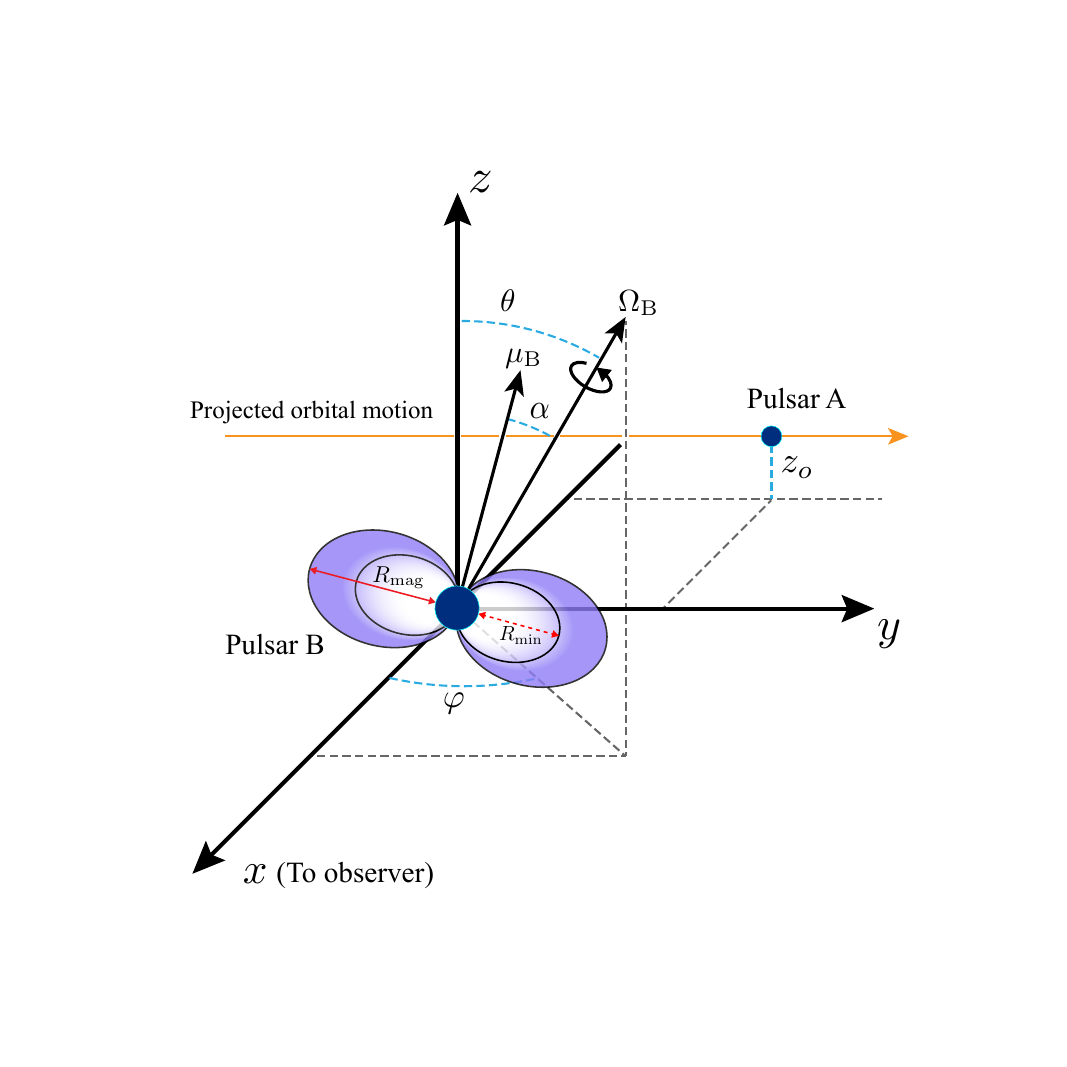}
    \caption{Diagram depicting the double pulsar geometry according to the eclipse light-curve model, adapted from Figure 1 of \citet{Breton2008}. 
    The minimum radial extent of the synchrotron absorbing plasma (the cooling radius) is represented by $R_{\min}$. Longitude and latitude of the spin axis of pulsar B ($\Omega_{\rm B}$) is given by the angles $\varphi$ and $\theta$ respectively. 
    The magnetic axis ($\mu_{\rm B}$) is offset from $\Omega_{\rm B}$ by angle $\alpha$. In our model we have neglected the deflection of A's radio signals in the gravitational field of B, which amounts to a change in the impact parameter at B of no more than about 600\, km \citep{Kramer2021b}.}
    \label{fig:0737_config}
\end{figure}


\subsection{Light-curve model}

Here we provide a summary of the light-curve model developed by \citet{Lyutikov2005b}.
Pulsar B is positioned at the centre of a Cartesian coordinate system where the $x$-axis points towards our line of sight to the system, the $z$-axis is in the plane of the sky as viewed from Earth, and the $y$-axis runs parallel to the apparent motion of pulsar A \citep{Breton2008}.
The motion of pulsar A is offset vertically from the origin by a constant value of $z = z_{0}$.
The direction of $\Omega_{\rm B}$ can be fully described by two angles: the spin-axis co-latitude ($\delta_{\rm B}$) with respect to the $z$-axis, and spin-axis longitude ($\varphi_{\rm so}$) with respect to the $x$-$y$-plane \citep{Damour1992}.
Ordinarily, $\delta_{\rm B}$ and $\varphi_{\rm so}$ would be related to the Cartesian coordinate system via
\begin{eqnarray}
    \cos\theta &=& \cos(90^{\circ} - i)\cos(180^\circ - \delta_{\rm B}) \nonumber\\
    && - \sin(90^{\circ} - i)\sin(180^\circ - \delta_{\rm B}) \cos\varphi_{\rm so},\\[2mm] 
    \sin\varphi &=& -\frac{\sin(180^\circ - \delta_{\rm B}) \sin\varphi_{\rm so}}{\sin\theta}.
\end{eqnarray}
However, the double pulsar is nearly edge-on from our perspective (so either $i = 89.35^{\circ} \pm 0.05^{\circ}$ or $90.65^{\circ} \pm 0.05^{\circ}$; \citealt{Kramer2021b}), meaning the $z$-axis is effectively (anti-)aligned with the total angular momentum vector. As a result, we can make the simplifying assumption that $\theta \approx 180^\circ - \delta$ and $\varphi \approx -\varphi_{\rm so}$. Note that in the coordinate frame of \cite{Breton2008}, where A moves in the positive $y$ direction during the eclipse, the orbital angular momentum is (almost) aligned with the negative $z$ direction, that is, pointing downwards in Figure~\ref{fig:0737_config}.

The total intensity models of pulsar A throughout the eclipses are computed from the synchrotron optical depth ($\tau$) of the plasma trapped within the closed-field region of pulsar B's magnetosphere. 
Following the prescription of \citet{Breton2008}, the optical depth at a given point during the eclipse is computed as
\begin{equation}\label{eqn:dpsr_tau}
    \tau = \frac{\mu}{\nu_{\rm GHz}^{-\ell}} \int_{-R_{\rm mag}}^{+R_{\rm mag}} \Bigg( \frac{B \sin\kappa}{B_{\rm mag}} \Bigg) \, d \Bigg( \frac{x}{R_{\rm mag}} \Bigg).
\end{equation}
The optical depth depends upon radio frequency ($\nu$) with a power-law index of $\ell$, $R_{\rm mag}$ is the truncation radius of B's magnetosphere, $B$ is the local magnetic field strength along the line of sight in units of $B_{\rm mag}$ (magnetic-field strength at $R_{\rm mag}$), $\kappa$ is the angle between the local magnetic field and our line of sight and $x$ is the radial position of pulsar A from our perspective in units of $R_{\rm mag}$.
Note that $R_{\rm mag}$ is not fit for directly but is inferred via the parameter $\xi$ that scales the size of the magnetosphere to 
the orbital distance between pulsars A and B \citep{Breton2009}.
The scaling parameter $\mu$ combines various parameters that describe the physical properties of B's magnetosphere as
\begin{equation}\label{eqn:dpsr_mu_scale}
    \mu = \frac{4.5 \times 10^{-6} \lambda_{\rm mag}}{N_{B}^{1/4}} \frac{k_{B} T_{e}}{10 m_{e} c^{2}},
\end{equation}
in which $\lambda_{\rm mag}$ is the pair plasma electron multiplicity, $N_{B}$ alters the size of the magnetosphere based on the impact of the wind from pulsar A, $k_{B}$ is Boltzmann's constant, $T_{e}$ and $m_{e}$ the electron temperature in the plasma and the electron mass, and $c$ is the vacuum speed of light.
Variations in each of these parameters, which are not practicable to be fit for individually, serve to alter the depth of the eclipses as the intensity of pulsar A is computed from the optical depth as $e^{-\tau}$.
In \citet{Lyutikov2005b} the frequency dependence of the optical depth is set at $\ell = 5/3$.
However, observations by \citet{Breton2012} showed the observed eclipse depth follows a much shallower frequency dependence than expected from the standard model.
We inferred a value of $\ell \sim 1/3$ from a multi-band fit to 16 subbands across a pair of eclipses detected with the UHF and L-band receivers.
We adopt this value throughout the remainder of this work.

The modulation pattern of the light curve depends strongly on the changing line-of-sight geometry of pulsar B as it rotates, which is modelled through the corresponding variations in both $B$ and $\kappa$ in Equation \ref{eqn:dpsr_tau}.
Both of these terms are related to the dipole unit vector magnetic polar angle ($\theta_{\mu}$) as
\begin{equation}
    \cos\theta_{\mu} = \frac{\hat{\mu} \cdot \mathbf{r}}{r},
\end{equation}
where $\mathbf{r} = \{x, y(t), z \}$, $r$ is the distance between pulsars A and B in spherical coordinates ($r = |\mathbf{r}| =  \sqrt{x^{2} + y^{2}(t) + z^{2}}$) and $\hat{\mu}$ is the dipole unit vector, the components of which are given by
\begin{eqnarray}\label{eqn:dpsr_mu}
    & \hat{\mu}_{x} = (\hat{\mu}_{x}^{\Omega} \cos\theta + \hat{\mu}_{z}^{\Omega} \sin\theta)\cos\varphi - \hat{\mu}_{y}^{\Omega}\sin\varphi, \nonumber\\
    & \hat{\mu}_{y} = (\hat{\mu}_{x}^{\Omega} \cos\theta + \hat{\mu}_{z}^{\Omega} \sin\theta)\sin\varphi + \hat{\mu}_{y}^{\Omega}\cos\varphi,\\
    & \hat{\mu}_{z} = \hat{\mu}_{z}^{\Omega} \cos\theta - \hat{\mu}_{x}^{\Omega} \sin\theta, \nonumber
\end{eqnarray}
with
\begin{eqnarray}\label{eqn:dpsr_mu_omga}
    & \hat{\mu}_{x}^{\Omega} = \sin\alpha\cos(\phi_{\rm B} + \Delta\phi_{\rm B}), \nonumber\\
    & \hat{\mu}_{y}^{\Omega} = \sin\alpha\sin(\phi_{\rm B} + \Delta\phi_{\rm B}),\\
    & \hat{\mu}_{z}^{\Omega} = \cos{\alpha}. \nonumber
\end{eqnarray}
Here, $\alpha$ is the magnetic inclination angle of pulsar B and $\phi_{\rm B}$ is the rotation phase of pulsar B, which is related to the spin-vector direction as $\phi_{\rm B} = \Omega_{\rm B}t = 2\pi t/P_{\rm B}$ where $P_{\rm B}$ is the spin period of pulsar B.
The parameter $\Delta\phi_{\rm B}$ accounts for the offset from $\phi_{\rm B} = 0$ at an assumed reference time.
The values of $B$ and $\kappa$ are computed at each step in $\mathbf{r}$ as
\begin{equation}
    B = \frac{\sqrt{1 + 3\cos^{2}\theta_{\mu}}}{r^{3}}\mu_{\rm B},
\end{equation}
and
\begin{equation}
    \cos\kappa = \frac{3 \cos\theta_{\mu}(x/r) - \hat{\mu}_{x}}{1 + 3 cos^{2}\theta_{\mu}},
\end{equation}
before being passed to Equation \ref{eqn:dpsr_tau} where they are integrated over $x$.

\subsection{Joint-fitting of eclipse pairs}\label{subsec:jointfit}

Unlike in \citet{Breton2008}, radio emission is not currently detected from pulsar B.
Without an a priori timing solution for pulsar B, we cannot predict the rotation-phase of pulsar B throughout our observations.
However, the $S/N$ of the eclipses detected by MeerKAT is sufficiently high that we can instead perform direct fitting of individual eclipses, where the period of pulsar B is held fixed at $P = 2.773$\,s and the rotation phase at the start of an eclipse timeseries included as a free parameter. 

Our initial attempts at fitting the eclipses independently of one another were hampered by `eclipse weather', epoch-to-epoch stochasticity in the eclipse envelope likely originating from fluctuations in the plasma content and radial extent of the closed magnetic field lines of pulsar B.
This phenomena resulted in excess scatter in the individual geometric constraints inferred from one eclipse to the next.
We were able to mitigate a substantial amount of this behaviour by performing joint fits to pairs of eclipses that were separated in time by only a single orbit.
This was conducted using a Gaussian likelihood function of the form
\begin{equation}
    \mathcal{L}(\mathbf{d} | \mathbf{\Theta}, \sigma_{Q}) = \prod_{i=1}^{2} \prod_{j=1}^{N} \frac{1}{\sqrt{2\pi\hat{\sigma}_{i,j}^{2}}} \exp \Bigg[-\frac{(d_{i,j} - \mu_{i,j}(\mathbf{\Theta}))^{2}}{2\hat{\sigma}_{i,j}} \Bigg],
\end{equation}
where $\mathbf{d}$ represents the input eclipse light curves, $\mu$ is the eclipse model, $\mathbf{\Theta}$ contains the model parameters, and $\hat{\sigma}_{i,j}^{2} = \sigma_{i,j}^{2} + \sigma_{Q,i}^{2}$ are the uncertainties on the light-curve fluxes added in quadrature with an additional error in quadrature parameter (EQUAD; $\sigma_{Q,i}$).
This extra uncertainty parameter accounts for both pulse-to-pulse flux variations of pulsar A and unaccounted systematic errors.
Our priors for the model parameters are summarised in Table \ref{tab:priors}.
Posterior samples for the model parameters were generated using {\sc bilby} \citep{Ashton2019} as a front-end to the {\sc PyMultiNest} sampler, a {\sc Python}-based implementation of the {\sc MultiNest} nested-sampling algorithm \citep{Buchner2014, Feroz2009}. 

From our initial test fits to the data, we found the posterior distributions for the cooling radius consistently displayed equal a posteriori support for values of $R_{\min} \lesssim 0.5$ before running up against the edge of the prior.
While this does not give us a precise radius at which the plasma becomes transparent to radio waves, we can at least constrain it to less than half the radius of pulsar B's truncated magnetosphere.
Given the template light-curves are identical for $R_{\min} \lesssim 0.5$, we opted to fix the value of $R_{\min} = 0.5$ for the remainder of our light-curve modelling.

\begin{table}[t]
\centering
\caption[]{Priors on the \citet{Lyutikov2005b} model parameters.}
\label{tab:priors}
$$ 
\begin{array}{lcc}
\hline
\noalign{\smallskip}
{\rm Parameter\,(symbol)} & {\rm Prior\,range} \\
\noalign{\smallskip}
\hline
\noalign{\smallskip}
{\rm Spin\,phase,\,eclipse\,1\,}(\Delta\phi_{\rm B,1})   & {\rm Uniform}(0, 1) \\
\noalign{\smallskip}
{\rm Spin\,phase,\,eclipse\,2\,}(\Delta\phi_{\rm B,2})   & {\rm Uniform}(0, 1) \\
\noalign{\smallskip}
{\rm Spin\,period\,}(P_{\rm B})                        & {\rm DeltaFunction}(2.773) \\
\noalign{\smallskip}
{\rm Magnetic\,inclination\,}(\alpha)                  & {\rm Uniform}(0, 90) \\
\noalign{\smallskip}
{\rm Spin\mbox{-}axis\,latitude\,}(\theta)                    & {\rm Uniform}(90, 180) \\
\noalign{\smallskip}
{\rm Spin\mbox{-}axis\,longitude\,}(\varphi)                  & {\rm Uniform}(-90, 90) \\
\noalign{\smallskip}
{\rm Cooling\,radius\,}(R_{\min})                      & {\rm DeltaFunction}(0.5) \\
\noalign{\smallskip}
{\rm Magnetosphere\,scaling\,parameter\,}(\mu)          & {\rm Uniform}(0.5, 10) \\
\noalign{\smallskip}
z\mbox{-}{\rm axis\,offset}\,(z_{0})                          & {\rm Uniform}(-1.0, 0.0) \\ 
\noalign{\smallskip}
{\rm Scaled\,magnetospheric\,extent\,}(\xi)             & {\rm Uniform}(0, 2) \\
\noalign{\smallskip}
{\rm EQUAD,\,eclipse 1\,}(\sigma_{Q,1})                & {\rm Uniform}(0, 1) \\
\noalign{\smallskip}
{\rm EQUAD,\,eclipse 2\,}(\sigma_{Q,2})                & {\rm Uniform}(0, 1) \\
\noalign{\smallskip}
\hline
\noalign{\smallskip}
\hline
\end{array}
$$ 
\end{table}

\subsection{Measuring the spin-precession rate of pulsar B}

The spin vector of a rotating body moving in the curved spacetime of a companion precesses about the total angular-momentum vector of the binary system.
In the case of the double pulsar, the total angular momentum vector is almost exactly aligned with the orbit-normal vector ($z$-axis in Figure \ref{fig:0737_config}).
As a result, the geodetic precession of pulsar B manifests as a time-varying change in the spin-axis longitude while the spin-axis latitude remains unchanged. 
Hence, the time-evolution of these two parameters can be written as
\begin{equation}
    \theta = \theta_{0},
\end{equation}
and
\begin{equation}\label{eqn:precession}
    \varphi(t) = \varphi_{0} - \Omega_{\rm SO}^{\rm B} (t - t_{0}),
\end{equation}
where $\theta_{0}$ and $\varphi_{0}$ are the co-latitude and longitude of B's spin axis at some reference epoch, $t_{0} = $\,MJD 59289.

As a phase-connected timing solution for pulsar B across our dataset is unavailable, we could not use the exact same simultaneous joint-fit method utilised by \citet{Breton2008} to infer the precession rate of pulsar B. 
Instead, we explored two alternative means for measuring the change in spin-axis longitude with time.
As with the joint-eclipse fits, both approaches made use of the {\sc PyMultiNest} sampler with {\sc bilby}.

\subsubsection{Hierarchical inference}\label{subsec:hierarch}

Our first method involved a second-stage hierarchical fit to the posterior samples we obtained for $\varphi$ at each observing epoch.
We used a hyper-likelihood function of the form
\begin{eqnarray}
    &&\mathcal{L}_{\rm tot}(\varphi_{0}, \Omega_{\rm SO}^{\rm B}, \sigma_{\varphi} | \varphi(t)) =  \\
    && \prod_{i=1}^{N_{e}} \frac{1}{n_{i}} \sum_{k}^{n_{i}} \exp \Bigg[ -\frac{(\varphi(t_{i})_{k} - \varphi_{0} + \Omega_{\rm SO}^{\rm B}\,(t_{i}-t_{0}))^{2}}{2\sigma_{\varphi}^{2}} \Bigg], \nonumber
\end{eqnarray}
where $N_{\rm e}$ is the number of epochs for which we have measured $\varphi$, $n_{i}$ is the total number of posterior samples from the $i$-th eclipse, and $\sigma_{\varphi}$ is a normalising factor that accounts for the variance in our measurements of $\varphi$.
Uniform priors were assumed for each of the hyper-parameters.

\subsubsection{Iterative inference}\label{subsec:iterate}

Our second approach to measuring the precession rate involved performing a global `iterative' fit to every eclipse light-curve simultaneously.
Instead of using a timing model for pulsar B to predict its rotation phase throughout our observations, as was done by \citet{Breton2008}, we used the median value for the rotation phase obtained from the initial eclipse-pair fits as our input for the phase of pulsar B at the beginning of each eclipse. 
This allowed us to avoid fitting for the rotation phase of the pulsar, thereby reducing the number of model parameters by $N_{\rm obs}$.
As with the eclipse-pair fits, we used a joint Gaussian likelihood function of the form
\begin{equation}
    \mathcal{L}_{\rm tot}(\mathbf{d} | \mathbf{\Theta}(t)) = \prod_{i=1}^{N_{\rm obs}} \prod_{j=1}^{N_{i}} \frac{1}{\sqrt{2\pi\hat{\sigma}_{i,j}^{2}}} \exp \Bigg[-\frac{(d_{i,j} - \mu_{i,j}(\mathbf{\Theta}(t)))^{2}}{2\hat{\sigma}_{i,j}} \Bigg],
\end{equation}
where the time dependence of $\mathbf{\Theta}$ comes from computing $\varphi$ at each eclipse epoch via Equation~\ref{eqn:precession}.
Aside from the rotation phase of pulsar B and the EQUAD used for each eclipse (which were also set to the median values from the eclipse-pair fits), we assumed uniform priors on all of the geometric parameters.
This technique allows us to directly measure for the values of $\Omega_{\rm SO}^{\rm B}$ and $\varphi_{0}$ when fitting the eclipses.

\section{Results and analysis}\label{sec:results}

\subsection{Geodetic precession and model limitations}\label{subsec:limitaions}

\begin{figure*}
    \centering
    \includegraphics[width=\linewidth]{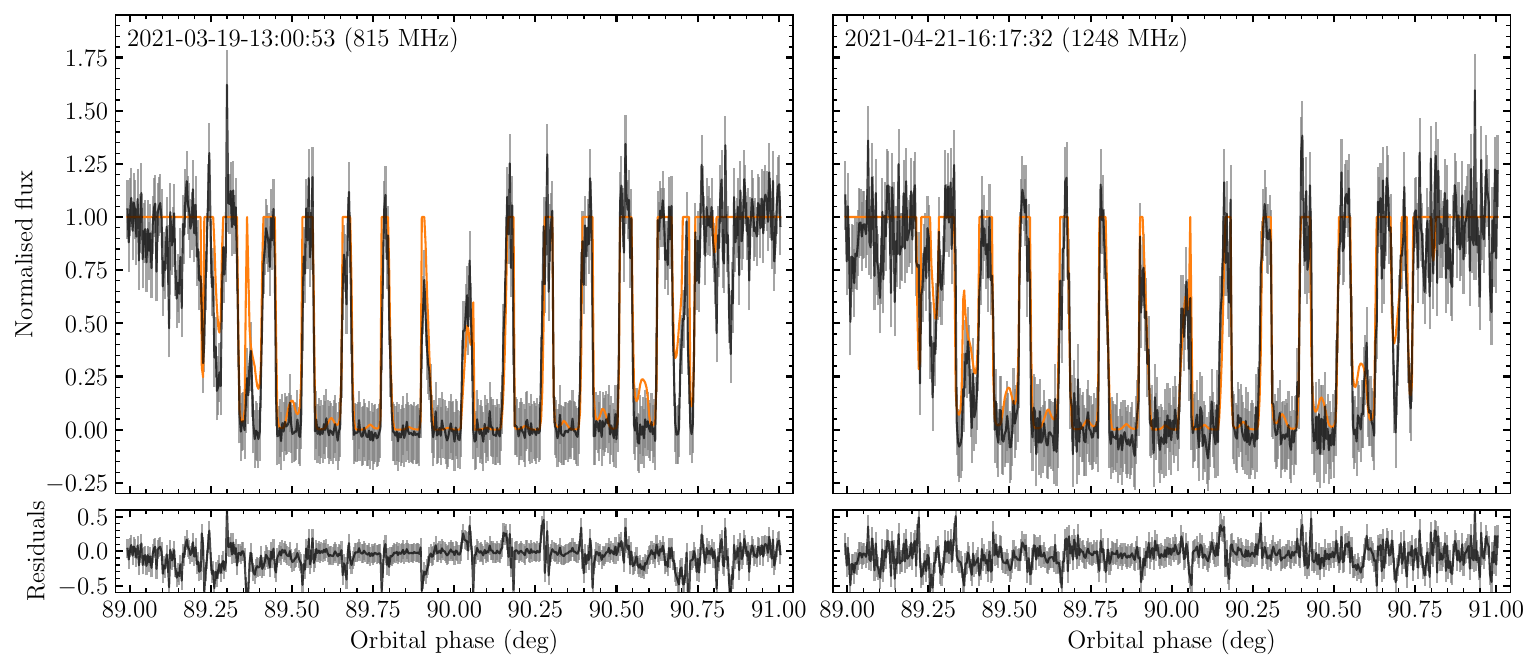}
    \caption{Two eclipses of pulsar A detected with MeerKAT using the UHF (left) and L-band (right) receiver systems. These light curves were generated after averaging over the full receiver bandpasses. We have inflated the flux uncertainties by the median recovered EQUAD for each eclipse. The orange traces correspond to the median a posteriori light-curve models recovered from fitting the observations. Bottom panels show the residuals after subtracting the median model.}
    \label{fig:0737_fits}
\end{figure*}

Our joint modelling of eclipse pairs were able to reproduce most of the observed peaks and troughs detected by MeerKAT at both L-band and UHF frequencies (see Figure \ref{fig:0737_fits}).
However, the eclipse model often failed to fully capture the morphology of the ingress and egress phases, where the effects of small variations in magnetosphere size or plasma density have the largest impact on the observed light curve.
Indeed, the median a posteriori models displayed in Figure~\ref{fig:0737_fits} show clear deviations from the data in both the ingress and egress phases.
This mismatch between the observed eclipse edges and the model was noted by \citet{Lyutikov2005b}, who found the best match occurred towards the centre of the eclipses.
The same effect was also seen by \citet{Breton2008}, which they compensated for by restricting their fits to only data between $-1.0^{\circ}/+0.75^{\circ}$ from superior conjunction.
In our case, the effects of these local distortions are amplified by the increased sensitivity of MeerKAT over the GBT.

\begin{figure*}
     \centering
     \begin{subfigure}[b]{0.49\textwidth}
         \vspace{0.3cm}
         \centering
         \includegraphics[width=\textwidth]{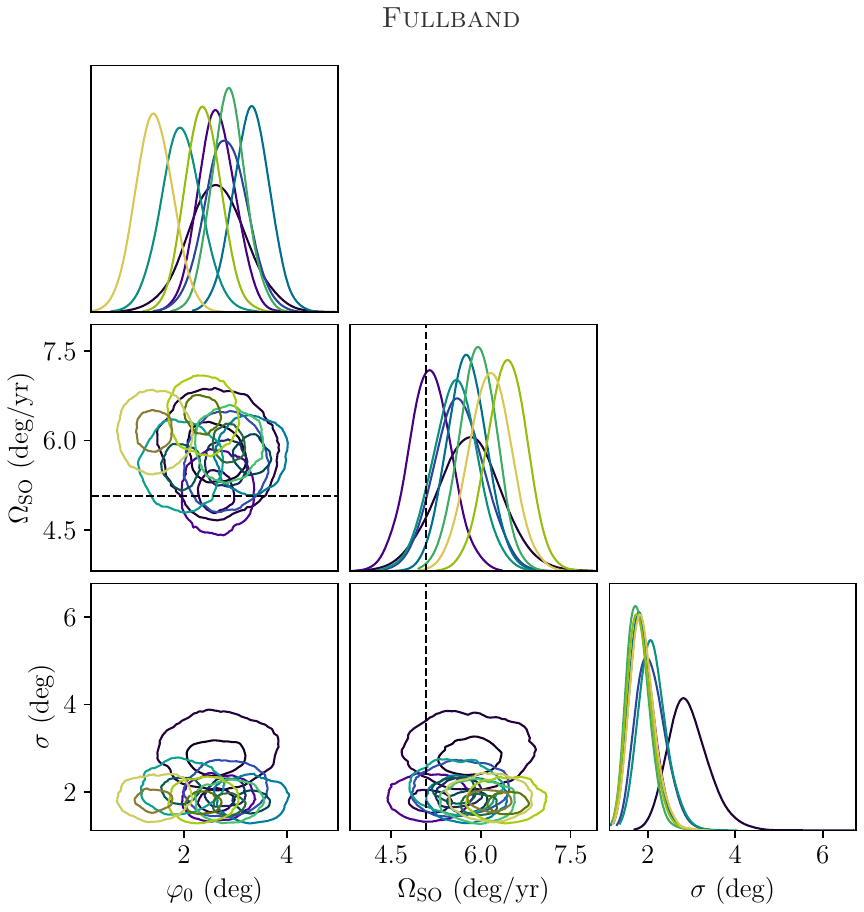}
     \end{subfigure}
     \begin{subfigure}[b]{0.49\textwidth}
         \vspace{0.3cm}
         \centering
         \includegraphics[width=\textwidth]{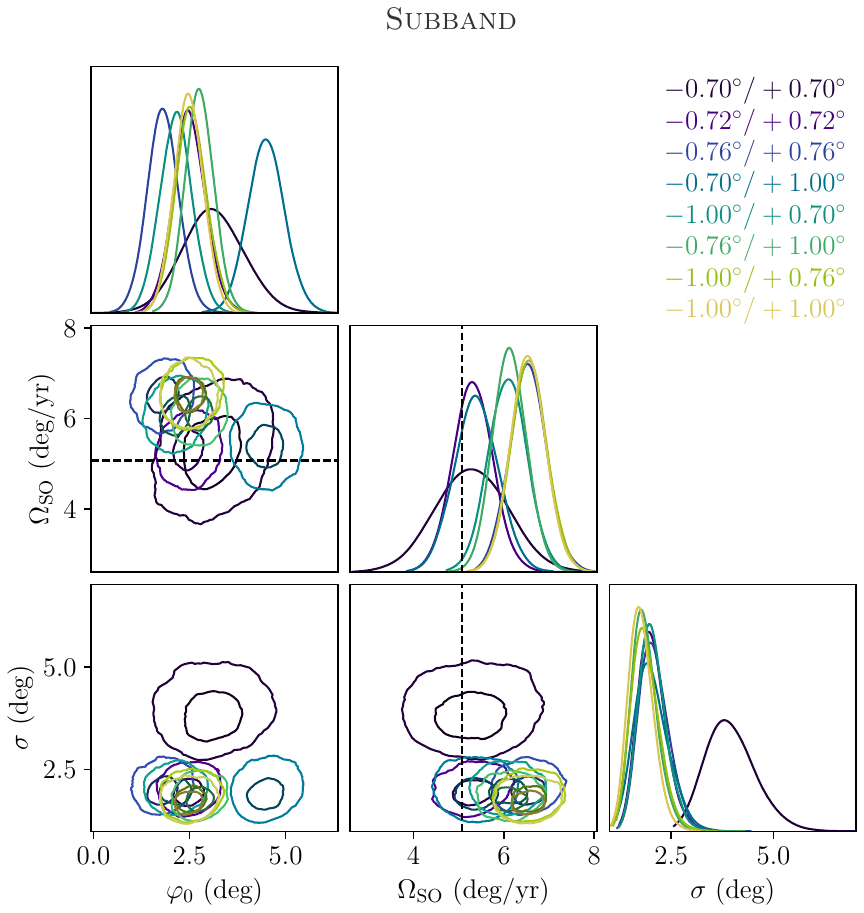}
     \end{subfigure}
    \caption{Comparison between hierarchical fits to different realisations of the light curves using the {\sc Fullband} and {\sc Subband} datasets. Contours in the two-dimensional posteriors represent the 68\% and 95\% confidence regions. Each colour represents measurements from eclipses that had different amounts of the ingress or egress phases removed. The dashed lines correspond to the predicted values of $\Omega_{\rm SO}^{\rm B}$ from general relativity.}
    \label{fig:comparison}
\end{figure*}

In Figure~\ref{fig:comparison}, we show that the recovered $\varphi_{0}$, precession rate and $\sigma_{\varphi}$ from hierarchical fits to both versions of the eclipse data vary depending on whether data is cut from the ingress and egress phases at different distances from superior conjunction.
Assuming the validity of general relativity, the predicted precession rate of pulsar B in rad\,s$^{-1}$ can be computed as \citep{Barker1975a}
\begin{equation}
    \Omega_{\rm SO}^{\rm B} = \frac{1}{2} \Bigg(\frac{G}{c^{3}}\Bigg)^{2/3} \Bigg(\frac{P_{b}}{2\pi}\Bigg)^{-5/3} \frac{m_{\rm A}(4m_{\rm B} + 3m_{\rm A})}{(1 - e^{2})(m_{\rm A} + m_{\rm B})^{4/3}},
\end{equation}
where $G$ is the gravitational constant, $c$ the vacuum speed of light, $P_{b}$ and $e$ are the orbital period eccentricity of the binary, $m_{\rm A}$ is the mass of pulsar A and $m_{\rm B}$ is the mass of pulsar B.
Using the pulsar masses and timing measurements from \citet{Kramer2021a}, we obtain a predicted precession rate of $\Omega_{\rm SO}^{\rm B} = 5.074005^{\circ} \pm 0.000003^{\circ}$\,yr$^{-1}$, which is shown as the dashed lines in Figure~\ref{fig:comparison}.
Comparing the different fits in Figure~\ref{fig:comparison}, it is clear that using the same restricted range as \citet{Breton2008} had little to no impact on the recovered model parameters when compared to runs that utilised the full eclipse light curves.
The vast majority of the restricted orbital phase ranges appear to result in measurements of $\Omega_{\rm SO}^{\rm B}$ that are biased towards larger values, appearing to be either marginally inconsistent with the GR prediction or consistent to within the 95--99.7\% confidence intervals.

The cause of this behaviour is apparent when comparing the light-curve shapes detected by MeerKAT between 2019 and 2023 to that detected by the GBT from 2003 and 2008.
From Figures~\ref{fig:0737_estack} and \ref{fig:0737_fits}, the light-curves appear relatively symmetric with the `partial' transparency window where some of the radiation from pulsar A leaks through the magnetosphere of pulsar B appearing around superior conjunction.
Notably, the transition points where the spacing between the transparency windows goes from half to a full rotation of pulsar B are restricted to the early-ingress and late-egress phases in our MeerKAT data.
The eclipses seen in the earlier GBT data (see Figure 2 of \citealt{McLaughlin2004} and Figure 3 of \citealt{Breton2008}) are substantially more asymmetric, with the half-to-full rotation transition taking place approximately one-third of the way through the eclipses, while the partial transparency windows appeared just prior to egress.
Accurately capturing the location of these transition points in our model fits is critical for recovering the correct spin-axis longitude over time.
The current locations of these features within the edges of the eclipse can account for the difficulties in measuring the pulsar geometry, as these regions are subject to the aforementioned local distortions in the magnetosphere of pulsar B, resulting in mismatches between the light-curve model and the data.
Both the $\pm 0.70^{\circ}$ and $\pm 0.72^{\circ}$ cuts result in hierarchical fits that are consistent with the GR-predicted value at the 68\% confidence interval with the {\sc Fullband} uncut dataset.

\begin{figure}
    \centering
    \includegraphics[width=\linewidth]{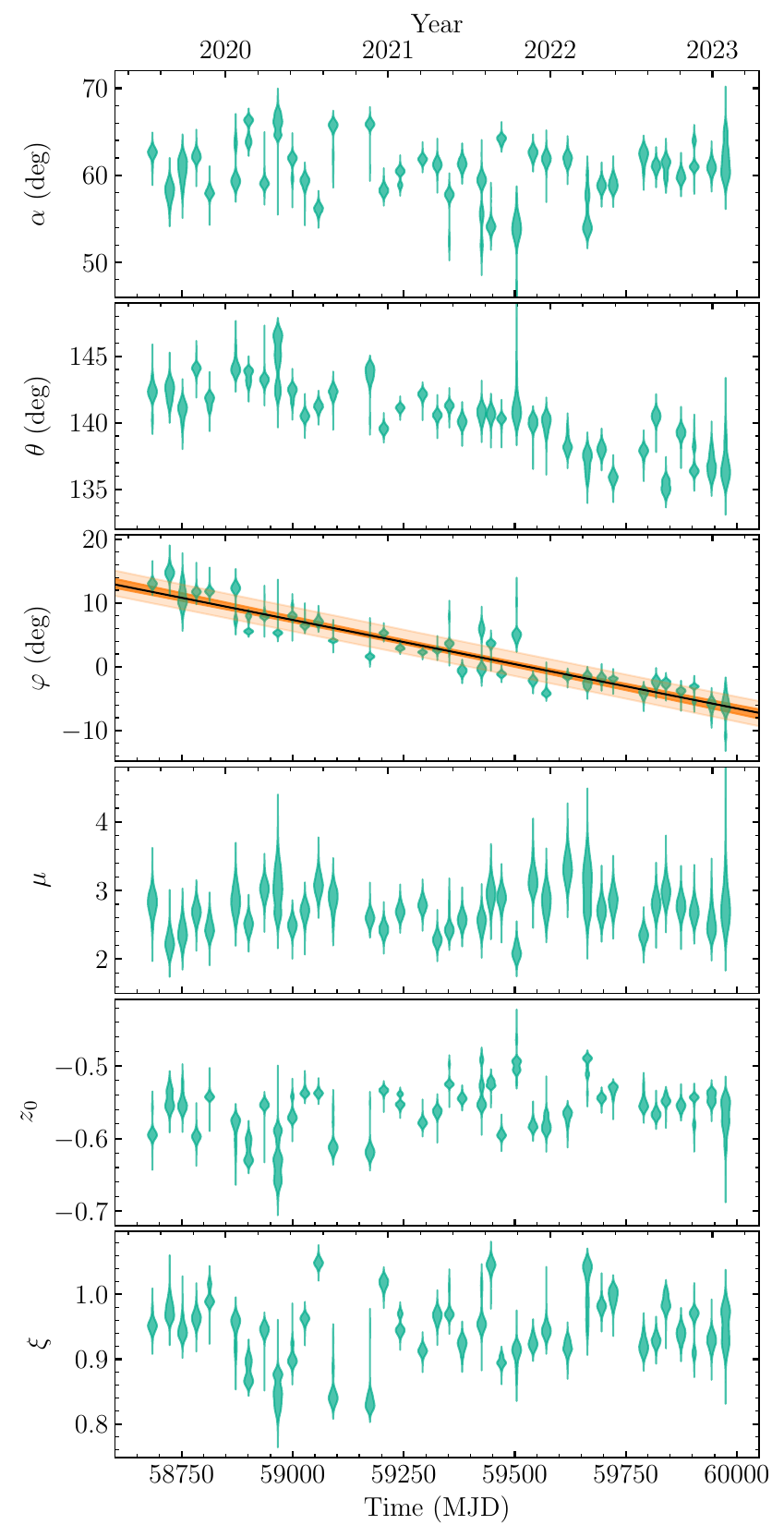}
    \caption{Evolution of our recovered geometric parameters for pulsar B over time measured using the {\sc Fullband} ($\pm 0.72^{\circ}$) data set. The GR-predicted evolution of $\varphi$ due to Geodetic precession is indicated by the black line. Dark orange shading indicates 68\% range covered by the precession rate inferred from our hierarchical fit. Light orange shading includes scatter in the values of $\varphi$ over time.}
    \label{fig:0737_geos}
\end{figure}

With these limitations in mind, for the remainder of this work we refer to results obtained using the {\sc Fullband} version of the eclipse data with the $\pm 0.72^{\circ}$ cuts unless otherwise specified.
In Figure~\ref{fig:0737_geos} we show the recovered geometric parameters for pulsar B from our fits to individual pairs of eclipses.
As expected, $\alpha$, $\mu$, $z_{0}$ and $\xi$ all remain at near constant values, while $\varphi$ displays a significant decrease from $13.0^{\circ} \pm 0.5^{\circ}$ to ${-6.2^{\circ}}^{+1.6^{\circ}}_{-1.4^{\circ}}$ as the spin axis of pulsar B precesses throughout the 3.5\,yr data set.
Notably, $\theta$ appears to show a slight downward trend which is not expected under our earlier assumption that $\theta = \theta_{0}$ in GR.
However, given the current unfavourable geometry and mismatch of the model to the data, this may simply be a result of time-dependent degeneracy between the evolving values of $\varphi$ and $\theta$.
We also over-plot the predicted precession rate of pulsar B in addition to that inferred from our hierarchical analysis of $\varphi(t)$ in Figure~\ref{fig:0737_geos}.
Our recovered precession rate of $\Omega_{\rm SO}^{\rm B} = {5.16^{\circ}}^{+0.32^{\circ}}_{-0.34^{\circ}}$\,yr$^{-1}$ agrees with the expected value of $5.074005^{\circ}$\,yr$^{-1}$ from GR to within $\sim$6.5\%.
For comparison, the corresponding precession rate obtained from the {\sc Subband} dataset with the same restricted fits to data within the ingress and egress returned $\Omega_{\rm SO}^{\rm B} = {5.31^{\circ}}^{+0.42^{\circ}}_{-0.41^{\circ}}$\,yr$^{-1}$.

In principle, the iterative method from Section~\ref{subsec:iterate} should provide a higher precision measurement of the precession rate as we average over more of the stochastic eclipse weather.
Performing this fit with the {\sc Fullband} dataset with $\pm 0.72^{\circ}$ cuts, we obtain $\Omega_{\rm SO}^{\rm B} = {5.27^{\circ}}^{+0.07^{\circ}}_{-0.08^{\circ}}$\,yr$^{-1}$. 
While the 68\% confidence intervals represent a $\sim$1.5\% uncertainty on $\Omega_{\rm SO}^{\rm B}$, the marginalised posterior distribution is only consistent with the GR-predicted value to within the 99.7\% confidence interval.
This is insufficient to use as a precision test of gravity and instead likely reflects the current difficulties in measuring the time-evolution of the spin-axis longitude with the existing light-curve model.

\subsection{Iterative eclipse fits}\label{subsec:iterfit}

\begin{figure*}
    \centering
    \includegraphics[width=0.95\linewidth]{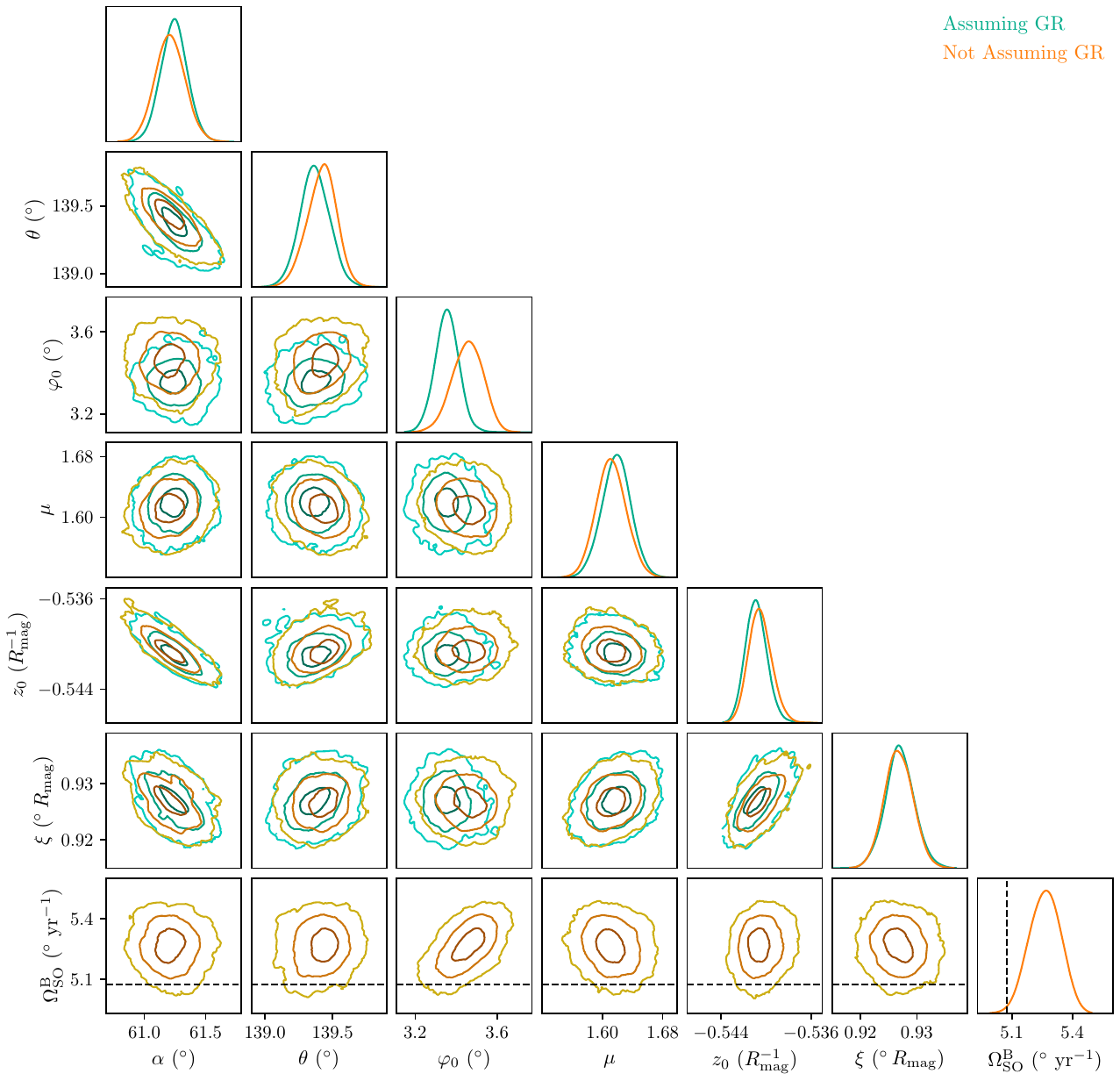}
    \caption{One- and two-dimensional posterior distributions of the eclipse light-curve parameters from iteratively fitting the {\sc Fullband} data with the $\pm 0.72^{\circ}$ cuts where the precession rate was either fixed to the GR predicted value (green) or allowed to be fit as a free parameter (orange). Contours indicate the 68, 95 and 99.7\% confidence intervals. Dashed black line is the predicted value of $\Omega_{\rm SO}^{\rm B}$.}
    \label{fig:big_fit}
\end{figure*}

While the current alignment of pulsar B is not conducive to performing rigorous, high-precision tests of spin-obit coupling, we can still obtain an accurate measurement of the overall system geometry under the assumption that GR is the correct theory of gravity.
Fixing the precession rate to the GR-predicted value of $5.074005^{\circ}$\,yr$^{-1}$, we used the iterative fitting approach with the {\sc Fullband} data restricted to fitting the $\pm 0.72^{\circ}$ region of the light curves to obtain high-precision measurements of the remaining geometric and magnetospheric parameters.
We present the recovered geometric and magnetospheric model parameters in Table~\ref{tab:params}.
In Figure~\ref{fig:big_fit} we present the one- and two-dimensional posterior distributions on the model parameters.
For comparison, we also show the same model parameters where $\Omega_{\rm SO}^{\rm B}$ was searched over as a free parameter. 

\begin{table}[t]
\centering
\caption[]{Recovered model parameters from the iterative fits to the {\sc Fullband} dataset using $\pm 0.72^{\circ}$ cuts with and without the assumption that GR is the correct theory of gravity. Uncertainties on each parameter represent the 68\% confidence intervals. The value of $\varphi_{0}$ is referenced to MJD~59289 (2021 March 16). In the last row we give the Bayesian evidences for the two models.}
\label{tab:params}
$$ 
\begin{array}{lcc}
\hline
\noalign{\smallskip}
{\rm Parameter\,(units)} & {\rm Assuming\,GR} & {\rm Not\,Assuming\,GR} \\
\noalign{\smallskip}
\hline
\noalign{\smallskip}
\alpha\,(^{\circ})                                 & 61.2 \pm 0.1 & 61.2 \pm 0.1 \\
\noalign{\smallskip}
\theta\,(^{\circ})                                 & 139.4 \pm 0.1 & 139.43^{+0.09}_{-0.11} \\
\noalign{\smallskip}
\varphi_{0}\,(^{\circ})                            & 3.36 \pm 0.06 & 3.47^{+0.07}_{-0.08} \\
\noalign{\smallskip}
\Omega_{\rm SO}^{\rm B} (^{\circ}\,{\rm yr}^{-1}) & -              & 5.27^{+0.07}_{-0.08} \\
\noalign{\smallskip}
\mu                                               & 1.62 \pm 0.02  & 1.61 \pm 0.02 \\
\noalign{\smallskip}
z_{0}\,(R_{\rm mag})                              & -0.5408 \pm 0.0009 & -0.5406^{+0.001}_{-0.0009} \\
\noalign{\smallskip}
\xi\,(^{\circ}\,R_{\rm mag}^{-1})                 & 0.927 \pm 0.002 & 0.927 \pm 0.002 \\
\noalign{\smallskip}
\ln \mathcal{Z}                                   &  16978.96 & 16979.01 \\
\noalign{\smallskip}
\hline
\noalign{\smallskip}
\hline
\end{array}
$$ 
\end{table}

We find only marginal differences in the recovered model parameters when we do or do not assume the GR-predicted precession rate. 
Indeed the negligible log-Bayes factor of $\ln\mathcal{B} = 0.05$ in favour of the `Not Assuming GR' model indicates the time-evolution of the eclipses is well described by pulsar B precessing at the GR-predicted rate.
However, our geometric and magnetospheric constraints are substantially different to those previously reported by \citet{Breton2008}
Notably, our $\alpha$ and $\theta$ differ by close to 10$^{\circ}$, while $\varphi_{0}$ (referenced to MJD~59289) is significantly different to the value predicted by \citet{Breton2008} for the same date.
These differences between eclipse modelling results likely originates from our framework fully sampling the model parameter space (except for $R_{\rm min}$), rather than setting $\mu$, $z_{0}$ and $\xi$ to fixed values.

We also tested whether evolution of the light-curve due to precession of pulsar B can break various covariances that exist between model parameters \citep{Breton2009}.
Symmetries exist between $\Delta\phi_{\rm B}$, $\alpha$, $\theta$, $\varphi$, and $z_{0}$, resulting in bimodal marginalised posterior distributions for the individual parameters and four modes in the two-dimensional posteriors with each other.
These symmetries appear as $\Delta\phi_{\rm} \rightarrow \Delta\phi_{\rm B} \pm 0.5$, $\{ \alpha, \theta, \varphi \} \rightarrow 180^{\circ} - \{ \alpha, \theta, \varphi \}$ and $-z_{0} \rightarrow +z_{0}$.
Since $\varphi$ varies with time due to pulsar B precessing, it is possible to break these symmetries by carrying out a simultaneous fit to several detected eclipses that are drawn from different points in our 3.5\,yr campaign.
We performed this joint fit to a sample of five eclipses using an expanded version of the approach detailed in Section~\ref{subsec:jointfit} with wider prior ranges on the aforementioned parameters of interest.
Our recovered posterior distributions for $\alpha$, $\theta$, $\varphi$, and $z_{0}$ were all well constrained and confined to a single mode peaking at the previously reported values in Table~\ref{tab:params}.
The only symmetries that were not broken via this approach were those between $\Delta\phi_{\rm B}$ and $\alpha$, where our recovered posterior distributions displayed two peaks separated by half a rotation of pulsar B for each $\Delta\phi_{\rm B}$ and $\alpha \sim 60^{\circ}$ or $120^{\circ}$.
Both come down to our lack of a priori knowledge on the rotation phase of pulsar B and could be resolved through independent timing of the pulsar via its radio pulses.

\section{Discussion}\label{sec:discussion}

\subsection{Testing theories of gravity}

\begin{figure}
    \centering
    \includegraphics[width=\linewidth]{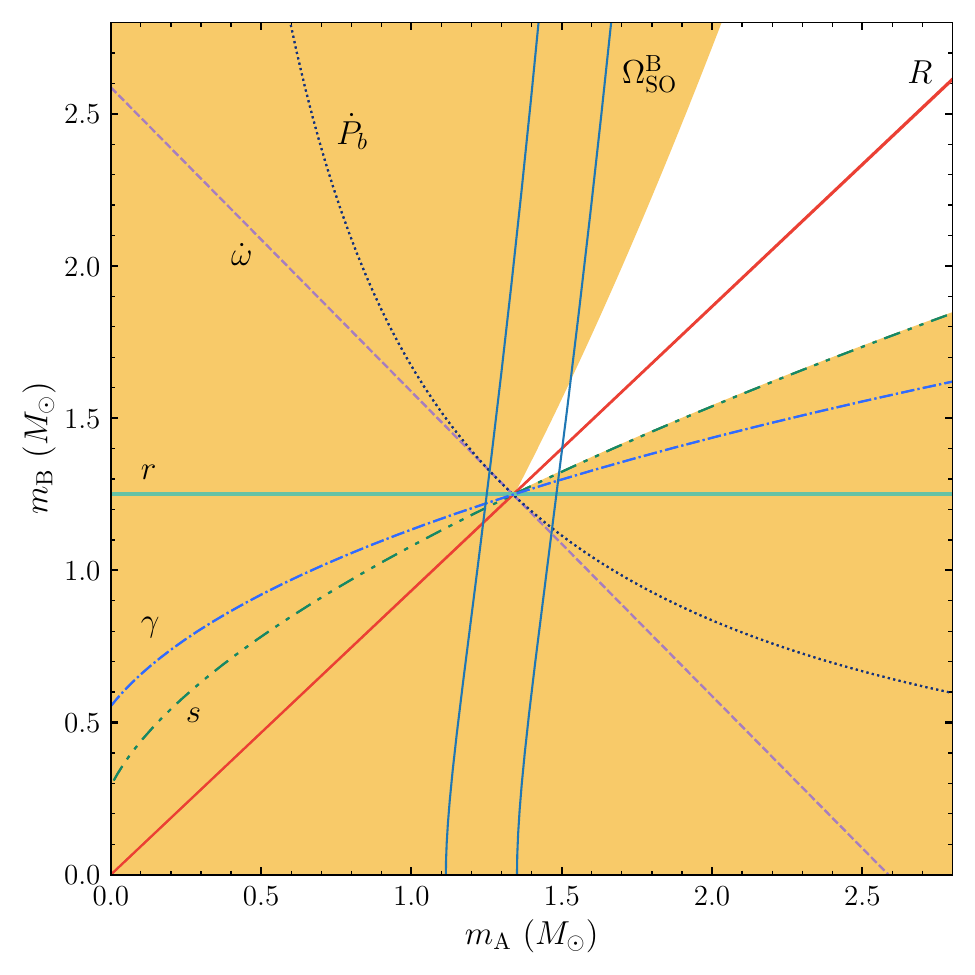}
    \caption{Mass-mass diagram illustrating current tests of general relativity with the double pulsar. Our improved measurement of pulsar B's spin-precession rate ($\Omega_{\rm SO}^{\rm B}$) is given by the (mostly) vertical blue lines where the separation between them indicates the 68\% uncertainty. Other individual constraints on the masses of pulsar A and B from \citet{Kramer2006, Kramer2021b} are shown by pairs of lines. These include the periastron precession rate ($\dot{\omega}$), shrinking of the orbital period due to gravitational radiation ($\dot{P}_{b}$), gravitational redshift ($\gamma$), Shapiro delay range ($r$) and shape ($s$), and the mass-ratio ($R$). The shaded region is forbidden by the individual mass functions of the two pulsars (i.e. $\sin i \leq 1$).}
    \label{fig:0737_mAmB}
\end{figure}

The geodetic precession rate can be used in combination with measurements of other relativistic effects to test GR and alternate theories of gravity.
In Figure \ref{fig:0737_mAmB}, we have plotted the inferred masses of pulsars A and B from $\Omega_{\rm SO}^{\rm B}$ alongside those from post-Keplerian parameters measured through precision timing of pulsar A \citep{Kramer2021b} and the mass-ratio \citep{Kramer2006} assuming the validity of GR.
A deviation of one or more pairs of lines away from a common range of pulsar masses would indicate an inconsistency with GR, assuming the impact of higher-order effects to measurements of the post-Keplerian parameters are either negligible or have been corrected for.
The pair of lines associated with the 68\% confidence interval for $\Omega_{\rm SO}^{\rm B}$ intersects the same common-point as the other post-Keplerian parameters, indicating that it is consistent with GR to our current measurement uncertainty.
This provides one of the five independent tests that are possible with the measurements displayed in this diagram.

In addition to the mass-mass comparison, the double pulsar is the only relativistic binary system in which direct constraints can be placed on the strong-field spin-orbit precession, as the measurement of $\Omega_{\rm SO}^{\rm B}$ can be combined with the orbital parameters inferred from timing both pulsars independently \citep{Breton2008, Kramer2009}.
Under the set of generic Lorentz-invariant relativistic theories introduced by \citet{Damour1992}, the geodetic precession rate can be reformulated as
\begin{equation}\label{eqn:dpsr_geo_test}
    \Omega_{\rm SO}^{\rm B} = \Bigg(\frac{P_{b}}{2\pi}\Bigg)^{-3} \frac{x_{\rm A} x_{\rm B}}{s^{2} (1 - e^{2})} \frac{c^{2} \sigma_{\rm so}}{\mathcal{G}},
\end{equation}
where $x_{\rm A}$ and $x_{\rm B}$ are the projected semi-major axes of pulsars A and B, $s = \sin i$ is the Shapiro-delay shape parameter, $\sigma_{\rm so}$ is the spin-orbit coupling constant and $\mathcal{G}$ is a generalised gravitational constant for the interaction between the two pulsars.
If GR is the correct theory of gravity, then we expect $\Big(\frac{c^{2}\sigma_{\rm so}}{G}\Big)_{\rm GR} = 2 + \frac{3}{2}\frac{m_{\rm A}}{m_{\rm B}} = 3.6076796 \pm 0.0000021$.
Substituting in our best value of $\Omega_{\rm SO}^{\rm B}$ and measurements for the post-Keplerian parameters from \citet{Kramer2021b} into Equation \ref{eqn:dpsr_geo_test}, we obtain $\Big(\frac{c^{2}\sigma_{\rm so}}{G}\Big) = 3.54 \pm 0.27$. 
Taking the ratio of the observed and predicted values, we find $\Big(\frac{c^{2}\sigma_{\rm so}}{G}\Big)_{\rm obs} / \Big(\frac{c^{2}\sigma_{\rm so}}{G}\Big)_{\rm GR} = (1 + 0.017) \pm 0.061$.
Hence, our measured spin-orbit coupling constant is consistent with the expectation from GR to within an uncertainty of $6.1$\%. 
The level of consistency between this test and the comparison between the predicted and observed precession rates from earlier is expected, as both are reliant on the same measured value.

These measurements of the geodetic precession rate of pulsar B and the associated test of spin-orbit coupling are encouraging, though the current unfavourable geometric alignment of pulsar B and systematic effects that arise from the model inconsistencies may introduce additional uncertainties on top of our reported statistical values.
Another source of systematic uncertainties we have not explored here is the impact the current lack of phase-coherent timing of pulsar B has on our measurements.
\citet{Breton2008} noted that variations in the input period of pulsar B alter both the expected rotational phase during the eclipses and duration of the transparency windows, which can lead to slightly faster or slower precession rates being inferred. 
As a result of these effects, the values presented here may have somewhat underestimated uncertainties. 

\subsection{System geometry}

Combined with precision timing of pulsar A over the past two decades \citep{Kramer2021b}, our eclipse modelling results provide the clearest picture to date of the three-dimensional geometry of PSR~J0737$-$3039A/B.
As noted in Section~\ref{subsec:iterfit}, the temporal evolution of the eclipse modulation pattern due to geodetic precession breaks several symmetries in the light-curve model.
From this we can unambiguously state that the spins of both pulsars are pointed out of the orbital plane of the system, with the spin axis of pulsar B being misaligned from the total angular momentum vector by $\delta_{\rm B} = 180^{\circ} - \theta = 40.6^{\circ} \pm 0.1^{\circ}$. 
This suggests pulsar B is rotating prograde in its orbit around pulsar A, consistent with beam-modelling results performed by \citet{Noutsos2020}.
Note that previous works incorrectly assumed that $\delta_{\rm B} \approx \theta$, where the seemingly large spin-axis offset was interpreted as arising from an off-centre supernova kick causing the newly born pulsar to tumble \citep{Farr2011}.
The corrected value of $\delta_{\rm B}$ suggests the supernova kick resulted in only a small offset in the final spin-axis away from the total angular momentum vector.
Combined with the precession rate, the angular offset between the spin axes of pulsars A and B can be computed as \citep{Perera2014},
\begin{equation}
    \cos(\Delta S(t)) = \cos(\delta_{\rm A})\cos(\delta_{\rm B}) + \sin(\delta_{\rm A})\sin(\delta_{\rm B})\cos(\varphi_{\rm SO}(t)),
\end{equation}
where $\delta_{\rm A}$ is the angle between the spin axis of pulsar A and the total angular momentum vector, $\delta_{\rm B} = 180^{\circ} - \theta$ and $\varphi_{\rm SO}(t)$ is computed from Equation~\ref{eqn:precession}.
Using the \citet{Perera2014} upper-limit of $\delta_{\rm A} \leq 2^{\circ}$, our measurements of $\theta$, $\varphi_{0}$ and the GR-predicted value of $\Omega_{\rm SO}^{\rm B}$, we obtain $\Delta S = 40^{\circ} \pm 2^{\circ}$ at our reference epoch of MJD~59289.
While our value is inconsistent with the $\Delta S = 138^{\circ} \pm 5^{\circ}$ obtained by \citet{Perera2014}, we note they had assumed that $\delta_{\rm B} \approx \theta$.
Correcting their measurement as $180^{\circ} - \Delta S$ returns a value of $42^{\circ} \pm 5^{\circ}$ which is in-line with what we obtained.

\begin{figure*}
    \centering
    \includegraphics[width=0.75\linewidth]{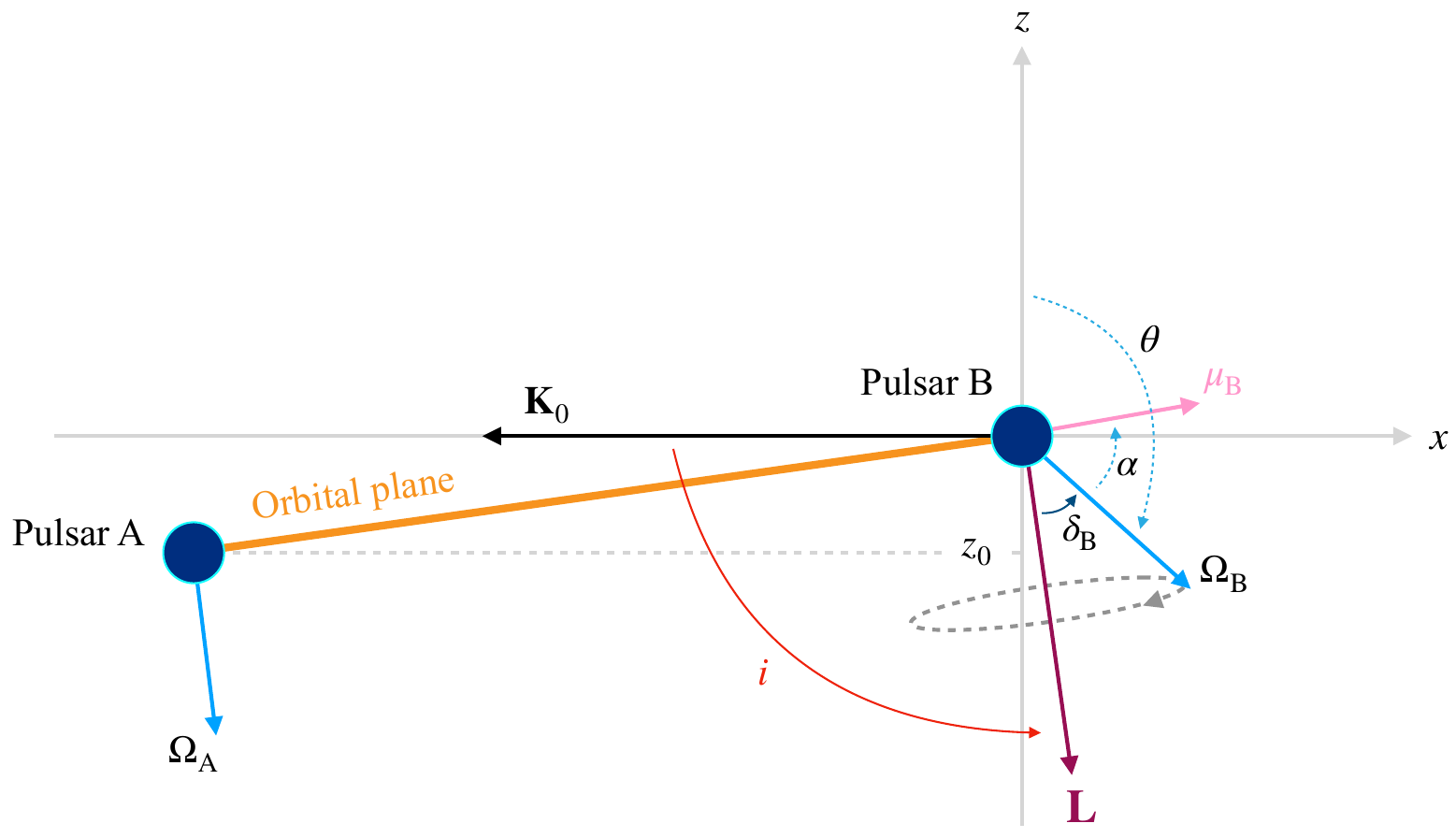}
    \caption{Diagram showing our inferred system geometry for the double pulsar using the same Cartesian coordinate system that is depicted in Figure~\ref{fig:0737_config}. The inclination angle ($i$) is defined as the angle between the orbital angular momentum vector ($\mathbf{L}$) and a vector pointing away from the observer ($\mathbf{K}_{0}$; \citealt{Damour1992}). We note that the offset of $i$ from $90^{\circ}$ is exaggerated by a factor of $\sim$20 in this image. All other angles are approximately correct.}
    \label{fig:0737_geom}
\end{figure*}

Measurements of the Shapiro delay shape parameter from timing pulsar A return two equally-likely solutions, where $i = 89.35^{\circ} \pm 0.05^{\circ}$ or $90.65^{\circ} \pm 0.05^{\circ}$ \citep{Kramer2021b}.
Attempts have been made to determine the `sense' of $i$ through measurements of the varying scintillation timescale of pulsar A throughout its orbit, with works by \citet{Ransom2004} and \citet{Rickett2014} finding $i = 88.7^{\circ} \pm 0.9^{\circ}$ and $88.1^{\circ} \pm 0.5^{\circ}$ respectively. 
A separate study of the scintillation properties of both pulsars by \citet{Coles2005} found $|i - 90^{\circ}| = 0.29^{\circ} \pm 0.14^{\circ}$.
Modelling of the double pulsar eclipses provides an independent means of determining the orbital inclination angle as the recovered values of $\theta$ and $\varphi_{0}$ depend on the sign of $z_{0}$.
Previously it was thought that symmetries between model parameters would make it difficult to determine the correct sign of $z_{0}$ \citep{Breton2009}.
However we demonstrated in Section~\ref{subsec:iterfit} that the precession of pulsar B effectively breaks this degeneracy, and only a single peak consistent with a negative value of $z_{0} = -0.5408 \pm 0.0009 \, R_{\rm mag}$ was recovered.
Assuming the small-angle approximation we can compute the orbital inclination by multiplying the $z$-axis offset by the magnetospheric extent, obtaining $i = 90^{\circ} - z_{0}\,\xi = 90.501^{\circ} \pm 0.001^{\circ}$. 
Note that $i$ in this case is measured using the `DT92/RVM' convention \citep{Damour1992}, where it corresponds to the angular offset of the orbital angular momentum ($\mathbf{L}$) from a vector pointing along our line from the Earth through the centre of mass ($\mathbf{K}_{0}$).
The full system geometry represented in the Cartesian coordinate system utilised by the eclipse model is illustrated in Figure~\ref{fig:0737_geom}.
At the apparent level of precision for which we measure $i$, it appears to be in a 3-$\sigma$ tension with the $i > 90^{\circ}$ value obtained from the Shapiro delay of pulsar A. 
But, as with the tests of GR, it is likely that the uncertainties on $z_{0}$ and $\xi$ are underestimated by a substantial factor due to unmodelled systematic effects.
In particular, both of these parameters should display some level of time-dependence as the projected distance between the pulsars vary due to their elliptical orbits undergoing periastron advance.
Implementation of these time-dependencies will be explored in future work.

Our recovered inclination sense disagrees with the aforementioned results from scintillation studies, where $i < 90^{\circ}$ is preferred.
Determining the orbital properties of the double pulsar via scintillometry is challenging due to substantial time-variations in the observed scintillation timescale, which \citet{Rickett2014} attempted to overcome by modelling the anisotropy of the turbulent screen along the line of sight.
However, it is possible that more complex geometric models are required to fully capture rapidly changing temporal variations in the scintillation pattern.
These improved models are currently under development using a combination of wide-bandwidth data collected with the MeerKAT UHF, L-band and S-band receivers (Askew et al., in prep.). 
Preliminary results indicate $i > 90^{\circ}$ is preferred.
We also note that an independent measure of the inclination angle from performing a rotating vector model fit to the linear polarisation of pulsar A under the assumption $\delta_{\rm A} \approx 0^{\circ}$ returned $i = 91.6^{\circ} \pm 0.1^{\circ}$ \citep{Kramer2021b}, albeit with uncertainties that only reflect the statistical errors in the measurement and do not take into account potential systematic effects.

\subsection{Implications for beam modelling and B's return}

Geodetic precession of pulsars in relativistic binaries results in our line of sight passing through different regions of their emission cone over time. 
This affords us the rare opportunity to map the active radio-emitting field lines above the polar cap.
The extent to which we can map a given pulsars beam depends on several factors: the shape of the active region, the beam filling factor, emission cone structure and pulsar geometry.
For pulsar B this is further complicated by the intense wind from pulsar A which distorts the active magnetic field lines around the polar cap of pulsar B and resulted in the pulsar only being detected in discrete `visibility windows' spread around different parts of the orbit \citep{Lyne2004, Lyutikov2005a, Lomiashvili2014}.
While the profiles detected in each window displayed slight differences in overall pulse shape, they did share the same secular evolution from having a clear single-peak to a double-peaked profile \citep{Burgay2005}. 
Attempts to model this evolution has resulted in two competing interpretations.
Early works suggested that pulsar B possesses a partially filled hollow cone, in which the radio-emitting region resembles an elongated horseshoe centred on the magnetic axis \citep{Perera2010, Perera2012, Lomiashvili2014}.
A more recent re-analyses of the profile evolution suggested pulsar B has a wedge-shaped beam consisting of two elongated Gaussian components that diverge towards the outskirts of the emission cone \citep{Noutsos2020}.
Both interpretations are strongly dependent on the assumed geometry of the pulsar, in particular the evolution of the angular offset between our line of sight and the magnetic axis ($\beta$) as the pulsar precesses and whether our line of sight was moving towards or away from the magnetic axis when pulsar B was visible.

\begin{figure}
    \centering
    \includegraphics[width=\linewidth]{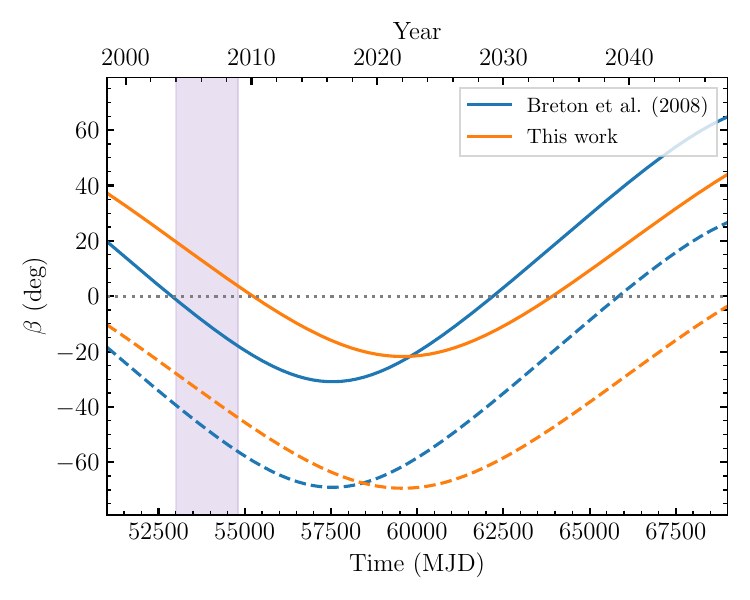}
    \caption{Predicted change in the impact parameter between the magnetic axis of pulsar B and our line of sight over time. Solid lines indicate the `active' beam that was detected in from 2003-2008, dashed lines correspond to the opposite magnetic pole. The vertical shaded region indicates the time-span over which radio pulses were detected from pulsar B.}
    \label{fig:0737B_return}
\end{figure}

The geometry and precession rate of pulsar B inferred from our eclipse modelling can be used to determine $\beta$ as per \citep{Breton2009},
\begin{equation}
    \cos(\pi - \zeta) = \sin(\theta)\cos(\varphi)\sin(i) + \cos(\theta)\cos(i),
\end{equation}
where $\beta = \zeta - \alpha$.
Using our measured geometry in Table~\ref{tab:params} and that reported by \citet{Breton2008}, we plot the resulting $\beta(t)$ curves from both geometries in Figure~\ref{fig:0737B_return}.
Tracks of $\beta(t)$ are shown for both the `active' magnetic pole that was visible from 2003-2008 and the antipodal pole from which we may detect radio pulses in the future.
Both sets of curves appear quite similar in terms of amplitude. 
There is however a substantial offset in the `phase' between the two assumed geometries, which comes from the $\sim$28$^{\circ}$ difference in $\varphi$ that are predicted for a given epoch.
Consequently, these two sets of results provide conflicting interpretations for the time-evolution of the pulse profile of pulsar B.
For the \citet{Breton2008} geometry, which was assumed to be correct in the beam-modelling of \citet{Noutsos2020}, the magnetic axis of pulsar B crossed our line of sight just prior to its discovery and the disappearance occurred when $\beta \sim -20^{\circ}$.
This was interpreted as emission cone of the pulsar missing our line of sight, with the corresponding value of $\beta$ setting a limit on the opening angle of the pulsar beam and motivated the aforementioned wedge-shaped beam map.
In contrast, the $\beta$ evolution corresponding to our inferred geometry is consistent with the elongated horseshoe beam model of \citet{Perera2010}, where the magnetic axis was initially at $\sim$20$^{\circ}$ when the pulsar was discovered and only crossed our line of sight in late-2009, after the radio pulses disappeared.
Despite the values of $\alpha$ and $\theta$ they recovered for the two orbital visibility windows varying in consistency with our measurements (Table~\ref{tab:params}), the $\varphi$ differs by only $\sim$4$^{\circ}$ when computed at the same reference epoch.
The values of $\alpha$ and $\theta$ obtained via an improved magnetospheric model for pulsar B in \citet{Perera2012} match our measurements extremely well.
However, the updated $\varphi$ is much closer to that of \citet{Breton2008}, differing from ours (at the same reference epoch) by $\sim$30$^{\circ}$.
This results in a large shift in the predicted pulse-shape evolution with time (compare Figure 17 of \citealt{Perera2010} and Figure 6 of \citealt{Perera2012}) and substantial differences in the predicted date that radio pulses will again be detected from pulsar B.
Given our improvements in modelling the eclipses over \citet{Breton2008}, we suggest that the beam shape of pulsar B most likely resembles the elliptical, partially filled hollow cone described by \citet{Perera2010}, or an inverted variant of the two-component wedge model where the separation between the two components increases as our line of sight cuts closer to the magnetic axis.

If the radio beam of pulsar B were symmetric about the magnetic axis then our geometry predicts we should have re-detected the pulsar mid-way through 2011, which did not occur.
Searches for radio pulses at the GBT and Parkes (`Murriyang') radio telescope over the past 15\,yrs have so far returned non-detections (R.~N. Manchester, private communication). 
This is interesting because little is known with certainty about the two-dimensional shape of a radio pulsar beam map (but see \citealt{Desvignes2019}).
As about half of all pulsars show multiple pulse components, the emission intensity  certainly varies along our line of sight (i.e. in a roughly longitudinal direction over the neutron star).
Yet, how the emission varies with latitude is mostly unexplored. 
The two-dimensional emission can be modelled as roughly circular; and shine completely, or only from  either cones or patches \citep[cf.~][]{lm88}.
Some models, on the other hand, predict fan-like beams \citep{dyks10}. 

The continued absence of pulsar B suggests that the lower half of the emission cone must be entirely devoid of active field lines. 
This is unusual for pulsars in general as we know emission arises on either side of the magnetic pole \citep{Johnston2023} and is not the case for any of the other precessing pulsars.
The number of other pulsar systems with beam maps to compare against is limited, with considerable variation in the underlying sources.  
PSR~B1913$+$16, on the one hand, is an old pulsar where recycling has diminished the magnetic field strength, and, arguably, influenced its configuration.
Its beam shape is consistent with either a hollow-cone or nested-hollow-cone depending on the chosen beam structure and viewing geometry \citep{Weisberg2002, Clifton2008}. In both cases the emission can be fit by a circularly symmetric conical beam. 
On the other hand, PSRs J1141$-$6145 and J1906$+$0746 are young and unrecycled. 
The radio beam of PSR~J1141$-$6145 appears to be largely filled  \citep{Manchester2010}; in some contrast to  the behaviour we infer for pulsar B. 
PSR~J1906$+$0746, however, combines both types of beam shape in a single source. 
This nearly orthogonal rotator allowed for the creation of beam maps from both poles \citep{Desvignes2019}. 
The evolution of its `interpulse' suggests that emission from this pole, the only one currently visible, is generally circular, and filled; similar to  PSR~J1141$-$6145.
Its `main pulse', however, which has since disappeared due to geodetic precession, consists of a narrow fan-like strip of emission that extends radially from the magnetic axis.
Irregular beam shapes are therefore possible and can help explain the continued absence of pulsar B.

The limited understanding of the shape of the pulsar B emission cone introduces substantial uncertainties when attempting to predict the return of its radio pulses.
\citet{Breton2009} and later \citet{Perera2012} predicted that the pulsar would re-enter the same range of $\beta$ values as 2003-2008 as soon as 2024 (see the solid blue trace in Figure~\ref{fig:0737B_return}).
However, our geometry predicts the same range will not be reached until early 2035, meaning the return of detectable radio pulses from pulsar B may be a decade later than initially predicted.
This is consistent with the earlier beam modelling of \citet{Perera2010}.
Similarly, radio emission from the opposite magnetic pole may not appear until sometime in the 2040's to 2050's as opposed to $\sim$2034 as predicted in \citet{Breton2009}.
In either case, the re-appearance of pulsar B, while requiring patience, will be an important input for understanding the beam shapes of non-recycled pulsars.

\section{Summary and conclusions}\label{sec:conclusion}

We presented the first results of our ongoing MeerKAT campaign to monitor the eclipses detected in the double pulsar system PSR~J0737$-$3039A/B.
Our Bayesian inference framework is capable of modelling the eclipse light-curve in the absence of a phase-coherent timing model of pulsar B, thereby allowing us to measure the pulsar geometry without a priori knowledge of the pulsar rotation phase.
However, the high sensitivity of MeerKAT, combined with the current geometry of pulsar B as viewed from Earth and limitations of the \citet{Lyutikov2005b} light-curve model presented significant challenges to obtaining a robust measurement of the geodetic precession rate of pulsar B.
Using a hierarchical Bayesian inference technique, we showed the recovered precession rate is strongly dependent on how much of the ingress and egress phases of the eclipses are included in the fits, thereby limiting the robustness of our current approach to measuring $\Omega_{\rm SO}^{\rm B}$ and the associated tests of general relativity.
With this in mind, we demonstrate that only fitting for the light-curve data within $\pm 0.72^{\circ}$ of superior conjunction returned a precession rate of $\Omega_{\rm SO}^{\rm B} = {5.16^{\circ}}^{+0.32^{\circ}}_{-0.34^{\circ}}$\,yr$^{-1}$, consistent with the expected value from GR to within 6.5\% uncertainty.
This in-turn provided an update to a theory-independent test of strong-field spin-orbit coupling, where our measurement of $\Big( \frac{c^{2}\sigma_{\rm so}}{G} \Big) = 3.54 \pm 0.27$ is consistent with GR at the 6.1\% level.
We showed the precession rate measurement could be improved via the use of an iterative framework to fit all of the observed eclipses simultaneously.
Yet the $\Omega_{\rm SO}^{\rm B}$ that we obtained via this approach is only consistent with GR to within the 99.7\% confidence interval. 
We suggest this is likely a result of the aforementioned unfavourable pulsar geometry and limitations of the existing eclipse model.

Our measurements of the system geometry are consistent with the spin axis of pulsar B being offset from the total angular momentum vector by $40.6 \pm 0.1^{\circ}$. 
This confirms that both pulsars A and B rotate in a prograde direction with respect to their orbital motion.
We also showed that our precession measurement broke several symmetries between model parameters, allowing us to determine the sense of the inclination angle $i = 90.501^{\circ} \pm 0.001^{\circ} > 90^{\circ}$. 
Upcoming scintillation measurements will confirm or rule out the presence of a tension with the $i < 90^{\circ}$ that is favoured in past studies.
Finally, we discussed our improved geometry in the context of past attempts to map the radio beam of pulsar B and provided an updated prediction for when radio pulses may again be detected.
Our inferred $\beta(t)$ suggest the radio beam likely resembles the elongated horseshoe shape presented in \citet{Perera2010}, and that the pulsar will not return to the same viewing geometry as 2008 until early 2035. 

Ultimately, many of the challenges faced in this work will be addressed through developing models of the pulsar B magnetosphere that better reflect the observed eclipse phenomenology. 
Improvements to account for frequency evolution across the wide fractional bandwidth of MeerKAT may also help reduce systematic uncertainties such that effective ToAs could be produced for pulsar B.
This would provide updated timing measurements of the double pulsar independent to pulsar A, as well as enable phase-coherent searches for radio pulses from pulsar B.
Construction of these models is particularly important given the near-future MeerKAT extension project and the eventual integration of MeerKAT into the Square Kilometer Array Mid-frequency telescope.
These will provide substantial increases in telescope gain, allowing us to resolve the eclipses in even greater detail than we present here.
Extending the baseline over which the precession rate is measured through continued monitoring with MeerKAT and linking with the long-running GBT monitoring campaign will help overcome issues pertaining to the unfavourable viewing geometry.
A combined MeerKAT+GBT dataset would return a precision measurement of $\Omega_{\rm SO}^{\rm B}$, enabling associated tests of gravity that are on-par with the measurements of other relativistic effects in this unique system.

\begin{acknowledgements}
The MeerKAT telescope is operated by the South African Radio Astronomy Observatory (SARAO), which is a facility of the National Research Foundation, an agency of the Department of Science and Innovation. 
SARAO acknowledges the ongoing advice and calibration of GPS systems by the National Metrology Institute of South Africa (NMISA) and the time space reference systems department of the Paris Observatory. 
PTUSE was developed with support from the Australian SKA Office and Swinburne University of Technology. 
This work made use of the OzSTAR national HPC facility at Swinburne University of Technology.
MeerTime data is housed on the OzSTAR supercomputer.
The OzSTAR program receives funding in part from the Astronomy National Collaborative Research Infrastructure Strategy (NCRIS) allocation provided by the Australian Government.
MK acknowledges significant support from the Max-Planck Society (MPG) and the MPIfR contribution to the PTUSE hardware. 
This work is supported by the Max-Planck Society as part of the `LEGACY' collaboration with the Chinese Academy of Sciences on low-frequency gravitational wave astronomy. 
RMS acknowledges support through Australian Research Council (ARC) Future Fellowship FT190100155. 
Part of the work was undertaken with support from the ARC Centre of Excellence for Gravitational Wave Discovery (CE1700004). 
RPB acknowledges support from the European Research Council (ERC) under the European Union’s Horizon 2020 research and innovation programme (grant agreement no. 715051; Spiders)
VVK acknowledges financial support from the ERC starting grant `COMPACT' (grant agreement number: 101078094).
JvL acknowledges funding from NWO Vici research programme `ARGO' (639.043.815).
MAM is supported by NSF Physcis Frontiers Center  award number 2020265.
Pulsar research at UBC is funded by an NSERC Discovery Grant and by the Canadian Institute for Advanced Research.
MEL thanks the MPIfR fundamental physics research group for their hospitality during a visit to work on this project.
This work has made use of NASA's Astrophysics Data System.
\end{acknowledgements}

%
%
\bibliographystyle{aa}
\bibliography{main.bib}

\end{document}